\documentclass[aps,prb,twocolumn,showpacs,floatfix,superscriptaddress,amssymb,amsmath]{revtex4}
\usepackage{graphicx}
\usepackage{dcolumn}% Align table columns on decimal point
\usepackage{bm}% bold math
\usepackage{amsmath}
\usepackage{fontenc}

\usepackage{color}
 \usepackage{wasysym}

\begin{document}
\title{Capacitively coupled double quantum dot system in the Kondo regime}

\author{Irisnei L. Ferreira}
\affiliation{Instituto de F\'i­sica - Universidade Federal de Uberl\^andia -
Uberl\^andia, MG  38400-902  - Brazil}
\author{P. A. Orellana}
\affiliation{Department of Physics, Universidad Cat\'olica del Norte, Casilla
1280, Antofagasta, Chile}
\author{G. B. Martins}
\email[Corresponding author: ]{martins@oakland.edu}
\affiliation{Department of Physics, Oakland University, Rochester, MI 48309,
USA}
\author{F. M. Souza}
\affiliation{Instituto de F\'i­sica - Universidade Federal de Uberl\^andia -
Uberl\^andia, MG  38400-902  - Brazil}
\author{E. Vernek}
\affiliation{Instituto de F\'i­sica - Universidade Federal de Uberl\^andia -
Uberl\^andia, MG  38400-902  - Brazil}

\date{\today}
\begin{abstract}
A detailed study of the low-temperature physics  of an interacting double
quantum dot system in a T-shape configuration is presented. Each quantum dot is modeled by a single Anderson impurity
and we include an inter-dot electron-electron interaction to account for capacitive
coupling that may arise due to the proximity of the quantum dots.
By employing a numerical renormalization group approach to a multi-impurity
Anderson model, we study the thermodynamical and transport properties of the
system in and out of the Kondo regime. We find that the two-stage-Kondo effect reported in
previous works is drastically affected by the inter-dot Coulomb repulsion. In 
particular, we find that the Kondo temperature for the second stage of the two-stage-Kondo 
effect increases exponentially with the inter-dot Coulomb repulsion, providing a 
possible path for its experimental observation. 
\end{abstract}

\maketitle
\section{Introduction}
Many-body electron-electron interaction is one of the most striking phenomena in
low dimension condensed matter systems. In this context, quantum
dots\cite{ScientAmerican.268.118} (QDs) have
played a prominent role in the  recent progress of theoretical
studies,\cite{PhysRevLett.82.3508,PhysRevLett.103.266806} as well
as experimental
realizations,\cite{Science.S.1998.540,Science.278.1792,J.Appl.Phys..101.09E102,
Nature.391.156,Phys.World.11.35}  as they offer a unique opportunity for
successful
measurement of many-body-related physical phenomena arising at low temperature
regimes. The relevance of electron-electron interactions in QD systems results
from the strong confinement of the electrons due to the reduced sizes
of typical
structures.\cite{PhysRevB.59.10165,PhysRevLett.82.390,PhysRevB.72.165349}
This interaction is responsible for several fascinating phenomena, e.g., Coulomb
blockade,\cite{PhysRevB.72.165349,PhysRevLett.81.5225} and Kondo
effect,\cite{Science.293.2221, Science.289.2105,Nature.391.156} leading to
characteristic behavior of the thermodynamical and transport properties,
which depend drastically on the number of QDs, as well as on their topological
configuration in the structure. In recent years, strong on-site interaction in
double\cite{NanoscaleResearchLetters.3.343,PhysRevB.73.035332,
PhysRevLett.103.266806,PhysRevB.65.241304,PhysicaE.34.608,
PhysRevB.78.085308,Nature.385.417,PhysRevB.75.195339,PhysRevB.71.075305,
Science.293.2221,PhysRevLett.97.096603} and
triple\cite{PhysRevB.76.085305,PhysRevLett.98.047203,PhysRevB.80.035119,
PhysRevB.78.075424,
PhysRevB.68.161303,PhysRevB.81.075126,PhysRevB.77.193306,PhysRevB.81.041310} QD
(DQD and TQD) structures have received  a great deal of attention when in the
Kondo regime. However, {\it on-site} electron-electron interaction does not
exhaust all the possibilities in multiple QD structures, as electrons can, due
to their proximity, interact with each other, even when located in different
QDs. It is important to note that recent advances in the lithography of 
lateral semiconductor QDs have allowed greater control over capacitively coupled
DQD systems in parallel or in
series.\cite{Appl.Phys.Lett..80.1818,Appl.Phys.Lett..82.1887,
PhysRevLett.98.056801,PhysRevLett.101.186804,
arXiv:1008.0071v1.cond-mat.mes-hall..} This makes it even more relevant to
better theoretically understand the effects of capacitive coupling over systems like the one studied in this paper. 
Only recently has this long-range interaction attracted more widespread attention of
experimental and theory groups
alike.\cite{PhysRevB.77.045309,Appl.Phys.Lett..68.1954,PhysRevB.50.4917,
PhysRevLett.102.136805,PhysRevB.82.085311,J.Phys.:Condens.Matter.18.6545}

The T-shape configuration, where two QDs are mutually
coupled via a tunneling  matrix element, while only one of them is coupled to
metallic contacts, has attracted considerable interest as it allows the study of the two stage Kondo (TSK) 
effect,\cite{PhysRevB.71.075305,PhysRevB.72.165309,PhysRevB.77.035120,
PhysRevB.81.115316} by fine control of the inter-dot tunnel coupling.  
This effect results from the progressive screening of the localized spin of the electron in each QD.
Due to different effective couplings of the electron residing in each QD to the
conduction band, these magnetic moments are screened at different
temperature scales, which allows for the definition of two distinct Kondo
temperatures. While the on-site Coulomb repulsion has received a great deal
of attention, the inter-dot Coulomb interaction in this particular system has
not, to the best of our knowledge, been considered yet. In this paper, we will explore
how the TSK effect changes when a capacitive coupling is included between the dots.
We will use the Numerical Renormalization Group (NRG) method\cite{RevModPhys.47.773,PhysRevB.21.1003,PhysRevB.21.1044}
to calculate the transport and thermodynamical properties of the DQD system.
We will show that this new ingredient is responsible for
dramatic changes in the low-temperature physics of the system.
In particular, as shown in Fig.~\ref{entsus},
the Kondo temperature for the second stage of the TSK effect increases exponentially
with the inter-dot Coulomb repulsion. This may have important consequences
to the experimental observability of this effect. 
The paper is divided as follows: 
In section II, we present the model and a brief description of the NRG method.
In section III, we present numerical results for thermodynamical quantities (entropy
and magnetic moment), and transport properties (conductance). Finally, in section IV,
we present our conclusions.
\begin{figure}[h]
\centerline{\resizebox{2.7in}{!}{
\includegraphics{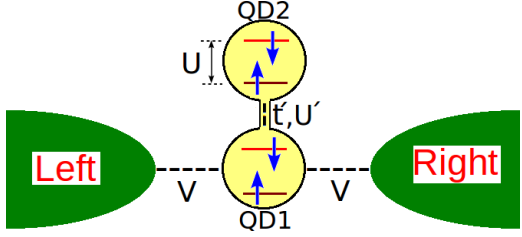}}}
\caption{\label{model}(color online) Schematic representation of the DQD system
being analyzed
in this work. Note the capacitive interaction, $U^\prime$,  between the dots.}
\end{figure}
\section{Theoretical model and numerical methods \label{sec-model}}
We study a system composed by two QDs (from now on referred to as QD1 and QD2)
coupled by a tunneling matrix element as well as by capacitive inter-dot Coulomb
repulsion. This system is described by the generalized Anderson Hamiltonian,
which can be written as 
% %
\begin{eqnarray}
 H=H_{imp}+H_{cb} + H_{hyb},
\end{eqnarray}
% %
 where $H_{imp}$ is the
Hamiltonian describing the QDs [which we define as  ``impurity region''],
$H_{cb}$ describes the conduction bands, and $H_{hyb}$ describes the coupling of
QD1
and the conduction bands (see Fig.~\ref{model}). More explicitly,
\begin{eqnarray}\label{H_imp}
 H_{imp}&=&\sum_{i=1,2\atop\sigma}\varepsilon_ic^\dagger_{i\sigma}c_{i\sigma}+
\sum_{i=1,2}U_i n_{i\uparrow}
n_{i\downarrow}+U^{\prime} n_1n_2\nonumber\\
&&+t^\prime\sum_{\sigma}\left[c^\dagger_{1\sigma}c_{2\sigma}+c^\dagger_{2\sigma}
c_{1\sigma}\right],
\end{eqnarray}
where the
operator $c^\dagger_{i\sigma}$ ($c_{i\sigma}$) creates (annihilates) an electron
in the $i$-th ($i=1,2$) QD, with energy $\varepsilon_i$, spin $\sigma$, 
$n_{i\sigma}=c^\dagger_{i\sigma}c_{i\sigma}$ is the number operator, and 
$n_i=\sum_\sigma n_{i\sigma}$. The second term in $H_{imp}$ corresponds to the
on-site Coulomb repulsion,  where, for simplicity we will take the intra-dot
interactions $U_1=U_2=U$ throughout this paper. The third term describes the
inter-dot Coulomb repulsion $U^{\prime}$  due to the proximity of the dots, and
the last term describes the coupling between the two dots, with tunneling matrix
element $t^{\prime}$. 
\begin{eqnarray}
 H_{cb}=\sum_{\ell k\sigma}\varepsilon_{\ell k} c^\dagger_{\ell k\sigma}c_{\ell
k\sigma},
\end{eqnarray}
where the operator $c^\dagger_{\ell k\sigma}$ ($c_{\ell k\sigma}$) creates
(annihilates) an electron with momentum $k$, energy $\varepsilon_{\ell k}$, and
spin $\sigma$ in the $\ell$-th lead ($\ell =L,R$). Finally,
\begin{eqnarray}
 H_{hyb}=\sum_{k\sigma}\left[V_{\ell k}c^\dagger_{1\sigma}c_{\ell k\sigma}+V_{\ell
k}^*c^\dagger
_{\ell k\sigma}c_{1\sigma}\right].
\end{eqnarray}
Notice that QD2
couples to the band {\it indirectly} through QD1.

For simplicity, we assume the hybridization coupling $V_{\ell k}=V$ to be real, 
independent of $k$, and the same for both leads. The conduction band is 
characterized by a constant density of states given by
$\rho_c(\omega)=\Theta(D-|\omega|)/2D$, where $D$ is the
half-bandwidth and $\Theta(x)$ is the standard Heaviside step function.  To
properly study the low-temperature physics of this setup we employ Wilson's
NRG\cite{RevModPhys.47.773,PhysRevB.21.1003,PhysRevB.21.1044} 
approach, which allows for a systematic assessment of the Kondo effect in
impurity systems. Within the NRG, we logarithmically discretize the  conduction band
and map it into a tridiagonal form, which corresponds to a semi-infinite chain
where the coupling between the sites has the form
\begin{eqnarray}
t_N=\frac{(1+\Lambda^{-1})(1-\Lambda^{-N-1})}{2\sqrt{1-\Lambda^{-2N-1}}\sqrt{
1-\Lambda^{-2N-3}}}~\Lambda^{-N/2},
\label{EQ.3.2}
\end{eqnarray}
where $\Lambda$ is the discretization parameter (all results shown here are for
$\Lambda=2.5$). The ``impurity" site ($N-1$) possesses  sixteen degrees of
freedom, corresponding to all the base-states necessary to fully describe the
quantum state of the two QDs, while the other $N\geq 0$ sites correspond to
single non-interacting sites of Wilson's chain. Denoting the base-states of
$H_{-1}=H_{imp}$ for the QDs as
\begin{eqnarray}
\mid\phi_i\rangle_{-1}\equiv\mid m_1,m_2\rangle_{-1}
\end{eqnarray}
where  $m_k=1,2,3,4$ corresponds respectively to $0, \uparrow, \downarrow$
or $\uparrow \downarrow$. In this basis the Hamiltonian $H_{-1}$ has matrix
elements
\begin{eqnarray}\label{Hmatrix}
[H_{-1}]_{ij}= _{-1}\hskip-0.1cm\langle \phi_i\mid H_{-1}\mid\phi_j\rangle_{-1}.
\end{eqnarray}
By diagonalizing  the matrix defined in Eq.~\ref{Hmatrix} we obtain a set of
sixteen eigenstates $\mid\Psi_i\rangle_{-1}$ with corresponding eigenenergies
$E_i^{(-1)}$. In terms of the base-states the eigenstates can be written as
\begin{eqnarray}
\mid \Psi_i\rangle_{-1}=\sum_jA_{ij}\mid\phi_j\rangle_{-1},
\end{eqnarray}
where $A_{ij}$ is the projection of the $i$-th eigenvector onto the
$j$-th base-state. Once we obtain the eigenstates, we calculate all the
necessary
matrix elements for the next iteration, after which  we add a new site $N=0$.
To describe the resulting system we enlarge the Hilbert space such that the new
basis is constructed performing all 64 possible combinations
\begin{eqnarray}
\mid\phi_i\rangle_0:=\mid m\rangle\otimes\mid \Psi_j\rangle_{-1},
\end{eqnarray}
where $m=1,\cdots, 4$ and $j=1,\cdots, 16$. This procedure is repeated until the
system has reached its strong-coupling fix point. When the dimension of the
Hilbert space becomes larger than $N_s$, where typically $N_s=2500$, it is
truncated by discarding the eigenstates corresponding to the
largest eigenenergies. At each iteration $N$ we keep the energy spectrum
$E^{(N)}_i$, together with the matrix elements necessary to calculate the
relevant physical quantities. This procedure allows us to calculate
thermodynamical properties, such as entropy $S$, magnetic
moment $\mu^2=k_BT\chi_{imp}$, spin-spin correlation $\langle
\textbf{S}_1\cdot\textbf{S}_2\rangle$, occupation numbers
$n_{i\sigma}\equiv\langle n_{i\sigma}\rangle$, etc, as well as
dynamical quantities, like density of states (DOS) and conductance.
For the entropy and magnetic moment it is usual to define the contribution from
the impurity as
$S_{imp}$ and $\mu^2_{imp}$. These quantities are generically written as
$X_{imp}=X-X_{0}$,
where $X_0$ is calculated in the absence of the impurity.
Within the canonical ensemble, as a function of temperature $T_N$, we can write,
\begin{eqnarray}
X(T_N)=\frac{1}{Z_N(T_N)}\sum_i {{_N}}\langle \Psi_i\mid \hat X\mid\Psi_i
\rangle_N e^{-\beta_NE^{(N)}_i},
\end{eqnarray}
where  $\beta_N=(T_N)^{-1}$, and
\begin{eqnarray}
 T_N=\frac{1}{2k_B\bar\beta}D(1+\Lambda^{-1})\Lambda^{-(N-1)/2}
\end{eqnarray}
 is a characteristic temperature associated to the $N$-th iteration, $\bar\beta$
is a real number of order 1, $\hat X$ is the operator associated to the quantity
$X$ and
\begin{eqnarray}
Z_N(T_N)=\sum_ie^{-\beta_NE^{(N)}_i}
\end{eqnarray}
is the canonical  partition function.

The conductance is calculated by the generalized Landauer formula
\begin{eqnarray}
G/G_0=-\sum_\sigma\int_{-\infty}^\infty{\tt
Im}[{\cal T}_\sigma(\omega)]\left[
\partial f(\omega)/\partial \omega\right]d \omega,
\end{eqnarray}
where ${\cal
T}_\sigma(\omega)=2\pi V^2\rho_c(\omega)G_{
11 }^{\sigma} (\omega)$, $G_0=(2e^2/h)$ and $G_{11}(\omega)$ is the Fourier
transform of the
full interacting double-time Green's function
\begin{eqnarray}
G^\sigma_{ii}(t,t^\prime)=-i\Theta(t-t^\prime)\Big\langle
\left[c_{i\sigma}(t),c^\dagger_{i\sigma}(t^\prime)\right] \Big\rangle,
\end{eqnarray}
which in the present case results to be spin independent.
The Green's function is calculated at frequency
$\omega_N=D(1+\Lambda^{-1})\Lambda^{-(N-1)/2}$
within NRG in a standard manner via Lehmann representation,
\begin{eqnarray}
G^\sigma_{ii}(\omega_N)&=&\sum_{nn^\prime}\frac{\mid {{_N}}\langle \Psi_n \mid
c_{i\sigma}\mid
\Psi_{n^\prime}\rangle_N\mid^2}{\omega_N-\left(E^{(N)}_n-E^{(N)}_{n^\prime}
\right)}\nonumber\\
&&\times\left(e^{-\beta E^{(N)}_{n}}+ e^{-\beta E^{(N)}_{n^\prime}}\right).
\end{eqnarray}
We also employ a logarithmic Gaussian broadening\cite{PhysRevB.64.045103} of the
discrete NRG spectrum
in order to obtain a smooth curve for the QDs DOS at arbitrary frequency
$\omega$
 \begin{eqnarray}
\rho^\sigma_{i}(\omega)=-\frac{1}{\pi}{\tt Im}G^\sigma_{ii}(\omega),
\end{eqnarray}
necessary to calculate the conductance at finite temperature.

\section{Numerical results \label{sec-results}}
In order to proceed with our  numerical analysis, let us set $D$, typically the
largest energy scale of the problem, as our energy unit ($D=1$).  We then choose 
for all calculations $U_1=U_2=U=5\times10^{-4}$
and $V=3.2\times10^{-3}$, so that $U/\pi\Gamma_1\approx 5$, where $\Gamma_1=2\pi V^2\rho_0$, 
and $\rho_0=1/(2D)$. We will study in detail the two cases where $U^\prime=0$ and $U^\prime=U$, 
and also the range $0 < U/U^{\prime} < 1$
The bare levels $\varepsilon_1$ and $\varepsilon_2$ will be controlled by the
same gate voltage ($V_g$), such that $\varepsilon_1=\varepsilon_2=V_g$. With the parameters set above,
and $V_g = -U/2$ (i.e., with the system at the particle-hole ({\it p-h})
symmetric point),
we can estimate the Kondo temperature for the single QD
as\cite{PhysRevB.21.1003}  
$T_K=\sqrt{\Gamma_1U}\exp(-\pi U/8\Gamma_1) \approx 2.89\times10^{-7}$ for QD1,
when QD2 is completely disconnected ($t^\prime=0$ and $U^\prime=0$).

\subsection{$U^\prime=0$ case}
\begin{figure}[h]
\centerline{\resizebox{3.4in}{!}{
\includegraphics{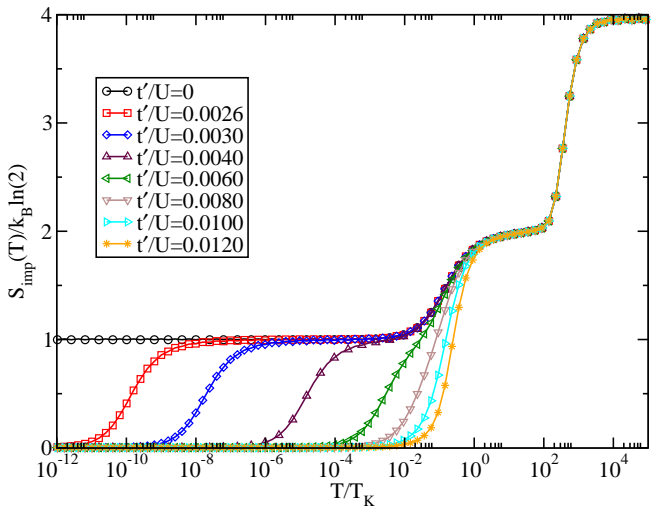}}}
\caption{\label{ent4}(color online) Temperature variation of the entropy for 
various values of $t^\prime$ (see legend).
The other parameters are $U_1=U_2=5\times10^{-4}$, $U^\prime=0$,
$\varepsilon_1=\varepsilon_2=-2.5\times10^{-4}$, and $V=3.2\times10^{-3}$. 
Note that the temperature in the horizontal axis is scaled by the Kondo temperature 
of QD1 (see text), which, for the parameters chosen, is $T_K=2.89\times10^{-7}$. Details 
for the different regimes are provided in the text.}
\end{figure}

\begin{figure}[h]
\vskip1cm
\centerline{\resizebox{3.4in}{!}{
\includegraphics{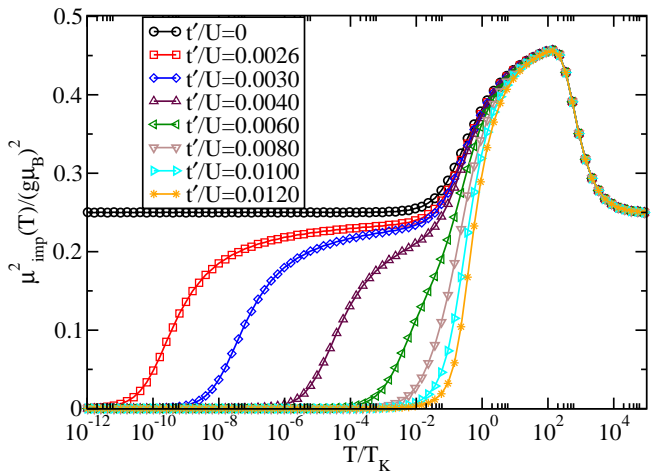}}}
\caption{\label{sus4}(color online) Magnetic moment $k_BT\chi$ as function of
temperature for various values of $t^\prime$. All parameters are the same as in Fig.~\ref{ent4}. 
}
\end{figure}

Although this case has been studied in great detail in
Ref.~\onlinecite{PhysRevB.71.075305}, we will present below some results that will 
help us understand the more complicated
situation at finite $U^\prime$.  In Figs.~\ref{ent4} and \ref{sus4}, we show results 
for the temperature dependence of the entropy and the square of the total magnetic moment 
of the DQD system, $S_{imp}/k_B \ln(2)$  and $\mu^2_{imp}/(g \mu_B)^2$, for  
$U^\prime=0$, $\varepsilon_1=\varepsilon_2=-U/2$ and  various
values of $t^\prime$, where $k_B$ is Boltzmann's constant and $\mu_B$ is the 
Bohr magneton. For the special case where $t^\prime=0$ [(black) \Circle curve], 
the DQD corresponds to the case where just QD1 is coupled to the conduction band
and QD2 is completely decoupled from the rest of the system. In this 
situation, as the temperature decreases, we find the following regimes: {\it (i)} for
$k_BT\sim D \gg U$ the DQD is in its free orbital (FO) regime; in this regime,  
the temperature is high enough to allow for  all the sixteen DQD states to
be populated. This results in an entropy $S_{imp}=k_B\ln(16)$ (see Fig.~\ref{ent4}) 
and a total square magnetic moment $\mu^2=2 \times (g\mu_B)^2/8$ (Fig.~\ref{sus4}), {\it (ii)}
for $k_BT_K<k_BT<\varepsilon_{1},\varepsilon_2$, thermally 
excited charge fluctuations are suppressed, and therefore the entropy decreases 
to $S_{imp}=k_B\ln(4)$, as only states with one electron in each QD are favored. 
This implies that $\mu^2$ increases, as double and unoccupied states in each QD 
are suppressed, and the DQD is in the so-called local moment (LM) regime.
As the temperature decreases further, and becomes lower than $T_K$, 
QD2 remains in its LM regime, while the other spin is progressively screened by the conduction
electrons due to the formation of the Kondo state, with the electrons in the
leads screening the spin of QD1. This regime is
characterized by the plateaus $S_{imp}=k_B\ln(2)$ and $\mu^2_{imp}=(g\mu_B)^2/4$.
These contributions arise just from the spin in QD2, as it is never Kondo screened when $t^\prime=0$.

For finite $t^\prime$, however, the behavior of the DQD for temperatures
below $T_K$ is quite different from the one just discussed above for $t^\prime=0$. 
For very small values of $t^\prime$, as shown in Ref.~\onlinecite{PhysRevB.71.075305}, 
the DQD system presents a TSK effect, where $both$ QD1 and QD2
spins are screened by the conduction band, but the screening of the spin 
in QD2 occurs at a much lower temperature than that at which the spin in QD1 is screened.
The second screening stage emerges at a characteristic temperature $T_K^\prime$ 
which depends mainly upon
$t^\prime$ and $T_K$ (the Kondo temperature for QD1) as 
\begin{eqnarray}\label{TK0}
T_K^\prime= a T_K\exp{(-b
T_K/J^\prime)},
\end{eqnarray}
where $a$ and $b$ are real positive numbers, with values approximately 1, and
$J^\prime=4t^{\prime 2}/U$ is an effective antiferromagnetic coupling
(between electrons in QD1 and QD2) favoring a local singlet state  that
competes with the regular Kondo energy scale $T_K$. As a result, a TSK effect is
expected for $J^\prime<T_K$. We can clearly see this behavior, for example, when 
$t^\prime/U=0.0030$ [(blue) $\Diamond$ curve] in Figs.~\ref{ent4} and \ref{sus4}. Note
that for this value of $t^\prime/U$ the second drop in the entropy (signaling the
screening of the electron in QD2) happens at $T_K^\prime \approx 10^{-17}$, with
$J^\prime\approx1.35\times 10^{-8}$ (for this value of $t^\prime$) being much
larger than $T_K^\prime$ (remember that $T_K \sim 2.89~10^{-7}$. As $t^\prime$ increases, and eventually
$J^\prime$ becomes larger than $T_K$, a strong singlet is formed
locally, destroying the TSK picture. A detailed discussion of this crossover can be 
found  in Ref.~\onlinecite{PhysRevB.71.075305}, here, instead, we focus on the
effect of the inter-dot Coulomb repulsion, as discussed in the next section.

\subsection{Finite $U^\prime$}
\subsubsection{Entropy and magnetic moment}

\begin{figure}[h]
\centerline{\resizebox{3.4in}{!}{
\includegraphics{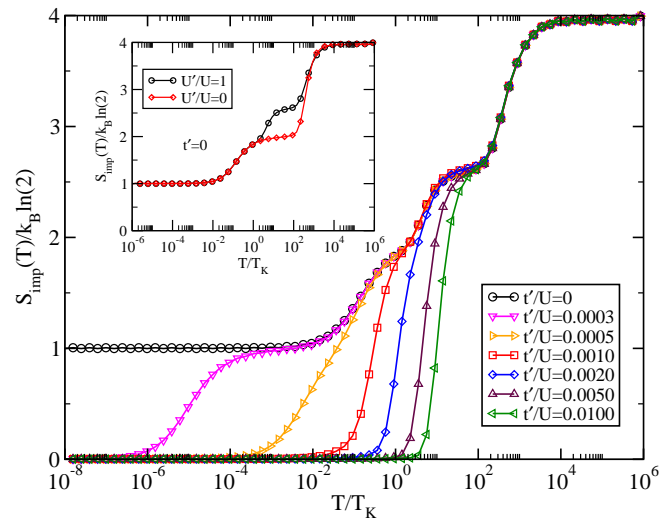}}}
\caption{\label{entfiniteu}(color online) Same as in Fig.~\ref{ent4}, 
but now for $U^\prime = U$ and different values of $t^\prime$ (see legend).
The Kondo temperature used to scale the horizontal axis is that obtained for 
$U^\prime = U$ and $t^\prime = 0$. The inset contains a comparison of 
$t^\prime=0$ results between $U^\prime=0$ [(black) $\Circle$ curve]
and $U^\prime=U$ [(red) $\Diamond$ curve], showing that the Kondo temperature of the first stage
(for $t^\prime=0$) does not depend on $U^\prime$.}
\end{figure}

Now, we turn on $U^\prime$ and for simplicity we choose $U^\prime=U$ 
(later on, we will analyze results for $0 < U^\prime < U$). 
The new features to note in the entropy (in relation to the results 
in Fig.~\ref{ent4}, for $U^\prime=0$) are {\it (i)} the very clear plateau (in Fig.~\ref{ent4}) 
at $S_{imp}/k_B \ln(2) = 2$ splits into two narrower plateaus in Fig.~\ref{entfiniteu}: one at 
$S_{imp}/k_B \ln(2) \approx 2.66$, and the one at $S_{imp}/k_B \ln(2) = 2$ becomes less well defined, 
more like a shoulder. The plateau at $\approx \ln(6)/\ln(2)=2.58$ (starting at 
$T \approx 10^2~T_K$) comes from the fact that now, as $U^\prime=U$, all 6 states with 2 electrons have very similar energies. 
As the temperature goes further down, the 4 states that can participate in a Kondo state in QD1 
(one electron in each QD) will have lower energy and the shoulder around 
$S_{imp}/k_B \ln(2) = 2$ will form (at $T \approx T_K$), {\it (ii)} to obtain the second stage of the TSK (signaled by 
a plateau at $S_{imp}/k_B \ln(2) = 1$, for finite $t^\prime$) one needs to go
to 
one order of magnitude smaller values of $t^\prime$, when compared to the results for 
$U^\prime = 0$ (compare the (purple) $\triangle$ curve in Fig.~\ref{ent4}, for $t^\prime=0.004$, 
with the (purple) $\triangle$ curve in Fig.~\ref{entfiniteu}, for $t^\prime=0.005$, where 
all traces of the second stage in the TSK effect have vanished), {\it (iii)} it is also 
apparent that, contrary to the $U^\prime = 0$ case, $T^\prime_K$ depends much more strongly 
on $t^\prime$. Indeed, it is clear that the temperature at which the entropy 
starts to decrease to zero becomes considerably higher as $t^\prime$ increases, 
conversely to what can be seen in Fig.~\ref{ent4}. It is important to note that 
the horizontal axis in Fig.~\ref{entfiniteu} is scaled by the Kondo temperature 
(for QD1) for $t^\prime=0$, estimated with the expression in the text above. In
the inset to Fig.~\ref{entfiniteu} 
we show that $T_K$ for $U^\prime=0$ and $U^\prime=U$ are equal. Indeed, as illustrated 
by the agreement between the entropy curves (for $T < T_K$) calculated with
$t^\prime=0$ [(black) $\Circle$ curve for $U^\prime=U$, and (red) $\Diamond$ curve for $U^\prime=0$]. 

\begin{figure}[h]
\centerline{\resizebox{3.4in}{!}{
\includegraphics{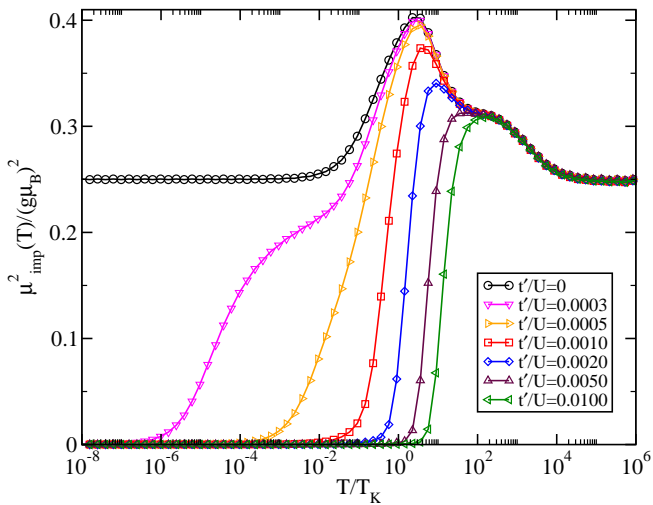}}}
\caption{\label{susfiniteu}(color online) Magnetic moment $k_BT\chi$ as function of
temperature for various values of $t^\prime$. All parameters are the same as in Fig.~\ref{entfiniteu}.}
\end{figure}

In relation to the magnetic susceptibility, a few differences are quite apparent when $U^\prime$ is turned on: 
{\it (i)} for the three smallest values of $t^\prime$, in Fig.~\ref{susfiniteu}, the peak in the susceptibility 
has moved down to $\approx T_K$, and the maximum value of the magnetic 
moment is now lower ($0.4$ for finite $U^\prime$, as compared to $\approx 0.45$ for 
$U^\prime = 0$). The decrease in the magnetic moment comes from 
the two extra states (two electrons in each dot) that are now part of this manifold of states, 
and which have zero magnetic moment; {\it (ii)} as $t^\prime$ increases, the peak moves up in temperature 
(up to $T = 10^2~T_K$) and the maximum value of the magnetic moment 
further decreases, down to $\mu^2_{imp}/(g\mu_B)^2 \approx 0.3$ for
$t^\prime=0.01$, 
{\it (iii)} for the three smaller values of $t^\prime$, the splitting of the single
plateau in the magnetic moment translates 
into a higher temperature shoulder (around $T/T_K=10^2$), which is not present 
for $U^\prime=0$.  This shoulder becomes also the maximum value of the magnetic
moment for $t^\prime=0.01$ (the largest $t^\prime$ results calculated).  

However, the most important differences coming from adding $U^\prime$ are related to the 
behavior of $T_K$ and $T_K^\prime$. First, the value of $T_K$ being used for the scaling of the 
horizontal axis is that obtained when $t^\prime=0$. Note from Figs.~\ref{ent4} and~\ref{sus4}, 
that $T_K$ is very weakly dependent on $t^\prime$ when $U^\prime = 0$. That is clearly not the case 
for $U^\prime=U$, where it can be seen a strong variation of $T_K$ and $T_K^\prime$ with $t^\prime$. 
In addition, more interestingly from an experimental point of view, it is clear that the 
ratio $T_K^\prime/T_K$ increases by a few orders of magnitude for $U^\prime=U$. 

\begin{figure}[h]
\centerline{\resizebox{3.4in}{!}{
\includegraphics{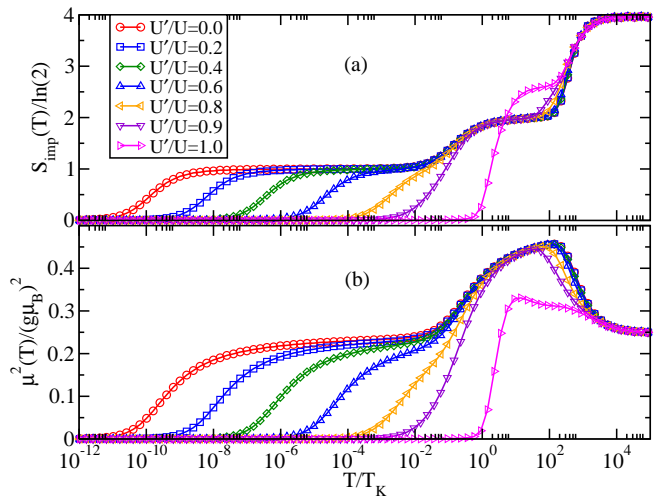}}}
\caption{\label{entsus}(color online) (a) Entropy $S_{imp}$ and (b) Magnetic moment $k_BT\chi$ as function of
temperature for various ratios of $U^\prime/U$ (between 0 and 1, see legend) 
for $t^\prime/U=0.0026$. To preserve the {\it p-h} symmetry for each curve
we set $V_g=-U/2+U^\prime$. All the other parameters are the same as in
Figs.~\ref{entfiniteu} and \ref{susfiniteu}. 
It is clear from the results that the Kondo temperature for the second stage of the TSK effect 
is greatly enhanced (by a few orders of magnitude) when $U^\prime$ increases.}
\end{figure}
\begin{figure}[h]
\centerline{\resizebox{3.4in}{!}{
\includegraphics{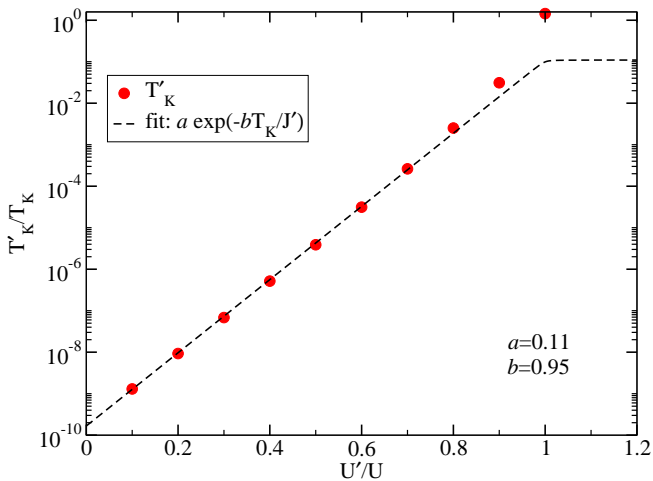}}}
\caption{\label{fit} (color online) Fitting (dashed line) of the
$T_K^{\prime}$ values [{red} dots] extracted from Fig.~\ref{entsus}, obtained by using
eqs.~\ref{TK0} and \ref{TK0new}. The values obtained (from the fitting) for the two free parameters {\it a}
and {\it b} are indicated in the figure.
}
\end{figure}

To analyze in more detail the increase of $T_K^\prime$ with $U^\prime$ 
we show, in Fig.~\ref{entsus}, results for entropy [panel (a)] and magnetic moment [panel (b)] which clearly 
indicate the strong increase in $T_K^\prime$ (by a few orders of magnitude) when 
the ratio $U^\prime/U$ varies from 0 to 1 for a fixed value of 
$t^\prime/U=0.0026$. This increase in $T_K^\prime$ can be understood by estimating the 
Kondo temperature $T_K^{\prime}$ for the second stage of the TSK, by using eq.~\ref{TK0}, 
and explicitly calculating $J^\prime$ for finite $U^{\prime}$: 
\begin{eqnarray}\label{TK0new}
J^{\prime} = -\frac{U-U^{\prime}-\sqrt{(U-U^{\prime})^2 +16t^{\prime2}}}{2}, 
\end{eqnarray}
indicating that, at least for small values of $U^\prime/U$ (where $J^{\prime} \sim 4t^{\prime2}/(U-U^{\prime})$, 
there should be an exponential increase of $T_K^\prime$ with $U^\prime/U$. 
This expectation is supported by the NRG results shown in Fig.~\ref{entsus}. One can go one 
step further and, substitute eq.~\ref{TK0new} into eq.~\ref{TK0}, and use the 
equation thus obtained to fit the values of $T_K^{\prime}$ that can be extracted from 
Fig.~\ref{entsus}. This fitting is shown in Fig.~\ref{fit}. Note that the only free parameters 
in the fitting shown in Fig.~\ref{TK0new} are `a' and `b' (values indicated in the figure). 
It can be clearly seen that fitting (at least up to $U^{\prime}/U \sim 0.7$ is very good 
(note that the vertical axis is in a logarithmic scale). Indeed, this is one of the principal results 
of this paper. Given the recent advances in the use of floating interdot capacitors (see Ref.~\onlinecite{Appl.Phys.Lett..80.1818}), 
values of $U^{\prime}/U$ for double dot systems have been steadily increasing, and may, in light of 
our results, offer the hope of observing the so far experimentally inaccessible energy scale $T_K^{\prime}$. 

\subsubsection{Zero-temperature case: QD occupation and conductance}

\begin{figure}[h]
\centerline{\resizebox{3.40in}{!}{
\includegraphics{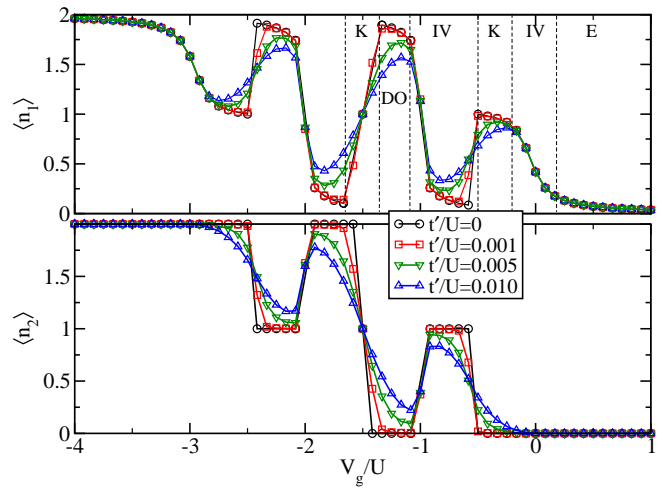}}}
\caption{\label{char6} (color online)
Total occupation $\langle n_1 \rangle$ (top)
and $\langle n_2\rangle$ (bottom)  as function of gate voltage
($V_g$) for  $t^\prime/U=0$ ($\Circle$), $t^\prime/U=1.0\times10^{-3}$
($\square$), $t^\prime/U=2\times10^{-3}$ ($\Diamond$) and
$t^\prime/U=1.0\times10^{-2}$ ({${\triangle}$}). Other parameters
are $t^\prime/U=5\times10^{-4}$ and
$V=3.2\times10^{-3}$. The regions delimited by the vertical lines
refer to QD1 states, namely, empty (E), intermediate valence (IV), Kondo (K) and
doubly occupied (DO) intermediate valence.}
\end{figure}

We first study how the occupation of the dots, $\langle
n_{i}\rangle\equiv\sum_\sigma\langle
n_{i\sigma}\rangle$, is modified due to inter-dot
Coulomb repulsion. In Fig.~\ref{char6}, we show
$\langle n_1\rangle$ (top) and $\langle
n_2\rangle$ (bottom) as function of $V_g$ for various values of $t^\prime$.
For  $t^\prime=0$ [(black) $\Circle$ curve], as $V_g\approx 0$ one can observe a smooth increase of $\langle n_1\rangle$, while
$\langle n_2\rangle$ remains zero all the way down to $V_g/U \approx -0.5$, where $\langle n_2\rangle$ jumps to $1$
and $\langle n_1\rangle$ abruptly decreases. The discontinuous increase of
$\langle n_2\rangle$ is due to the sharp peak in the local density of states
(LDOS) 
$\rho_{2}$ of level $\epsilon_2$, which is not broadened for $t^\prime=0$.
On the other hand, $\langle n_1\rangle$ increases smoothly, as $\epsilon_1$ is broadened by $\Gamma_1$.
The  value of $V_g$ for which the first  discontinuity of $\langle n_2\rangle$
occurs 
(expected to be at $V_g\approx 0$, when $U^\prime=0$)
depends on the additional capacitive energy necessary to put electrons in different dots,
due to the inter-dot  Coulomb repulsion $U^\prime$.

As discussed by three of the current authors in a previous work,\cite{PhysRevB.83.125404}
this gate-voltage-dependent charge
oscillation can be understood as a competition between the Kondo and
Intermediate Valence (IV) regimes: level $\epsilon_2$, when
$t^\prime=0$, acts as a
dark state, whose charge occupation, which can only be an integer, adds a step function
of height $U^\prime$ to the gate potential $V_g$ in QD1 (depending on $\langle
n_2\rangle$ being 0 or 1). The value of $V_g$ for which the transition of
$\langle n_2\rangle$ between 0 and 1 (and vice versa) occurs depends on
$\Gamma_1$ and reflects which many-body regime, Kondo or IV,
better optimizes the energy of QD1. The dark state $\varepsilon_2$ acts like a
switch between the two regimes and may have applications in quantum
computation.\cite{PhysRevA.82.042322,PhysRevB.83.125404} For example, when the
first discontinuity occurs, 
for $V_g/U \approx -0.5$, QD1 is in a Kondo state, however, further decrease of $V_g$ 
makes it more favorable for QD1 to be in an IV state, which can be accomplished 
by charging QD2 (by exactly one electron), this, due to the capacitive coupling, 
increases the effective gate potential of QD1, discharging it, and bringing it 
back to the IV regime.
As $V_g/U$ further decreases below $-0.5$, QD1 starts to transition from an IV to a Kondo 
state; when $V_g/U \approx -1$, again the IV state for QD1 is more favorable; this regime 
can now be achieved by completely avoiding the Kondo regime through the
discontinuous charging of QD1 by one extra electron. Total charge is kept constant by
discharging QD2 completely 
($\langle n_2\rangle=0$). Finally, further decrease of $V_g$, 
bringing it close to the  {\it p-h} symmetric point
[$V_g=-(U/2+U^\prime)=-3U/2$], increases the charge in QD1 to almost 2 electrons; 
at this point, the Kondo state in QD1 is more favorable, and a new discharging and charging occurs of QD1 and QD2, 
respectively, after which, each QD hosts one electron. 
These gate-voltage-dependent occupancy oscillations are clearly observed for all
curves in Fig.~\ref{char6}, 
with the difference that, for finite $t^{\prime}$, there are no discontinuities. 
Indeed, for finite $t^\prime$, discontinuities in the $\langle n_i\rangle$ are no longer 
observed, we rather notice that the jumps  are smoothened out as $t^\prime$
increases. This continuous charging of QD2 now results from the broadening
$\Gamma_2$ of the local bare (and also the many-body) level at QD2. For the
non-interacting case it is easily shown that $\Gamma_2\propto t^{\prime
2}\Gamma_1$. It is important to note the differences and similarities 
between this model and the one studied in Ref.~\onlinecite{PhysRevB.83.125404}: in the latter, 
one has a two channel system, where the dark state, for small values of $t_-$, 
acquires a finite broadening and smoothen out the discontinuities seen 
in the $\langle n_1 \rangle$ and $\langle n_2\rangle$ curves in Fig.~\ref{char6}. In the model 
being studied here, a similar process occurs: the dark state in QD2 acquires a broadening 
through its connection $t^{\prime}$ to the {\it single} conduction channel {\it through} 
QD1. Nonetheless, through a comparison of Fig.~\ref{char6} in this work with Fig.~4 in 
Ref.~\onlinecite{PhysRevB.83.125404} [panels (c) and (d)], one can see that qualitatively the results
are very similar, indicating that the basic processes determining the gate
voltage dependent occupancy oscillations are the same. 

\begin{figure}[h]
\vskip1.5cm
\centerline{\resizebox{2.80in}{!}{
\includegraphics{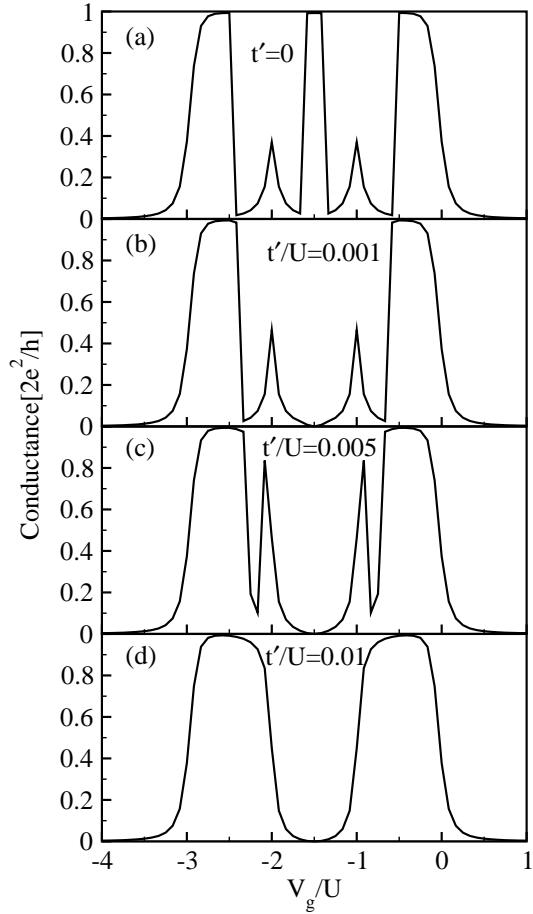}}}
\caption{\label{cond1}(color online) Zero-temperature  conductance as a function
of $V_g$ for the same parameters as in Fig.~\ref{char6}. The occupancy discontinuities 
analyzed in Fig.~\ref{char6} are associated to the discontinuities seen here for the conductance. 
See text for a full description of the variation of the peak structures with increasing 
$t^\prime$: (a) $t^\prime/U=0$, (b) $t^\prime/U=10^{-3}$, (c) $t^\prime/U=5 \times 10^{-3}$, and (d) $t^\prime/U=10^{-2}$.}
\end{figure}

The strong gate-dependent variations in $\langle n_i \rangle$ ($i=1,~2$) 
({\it caused by the capacitive coupling}) are expected to have 
dramatic influence in the conductance of the system, specially  in the Kondo
regime. For $U^\prime=0$, and in the special case where $t^\prime=0$, 
QD1 will be in a Kondo state for all $V_g$ values in the interval
$[-U,0]$. 
For finite $U^\prime$, however, the charging and discharging of the
QDs moves QD1 from a Kondo to an IV regime, and vice versa. 
For the case of $t^\prime=0$ [(black) $\Circle$ curve in Fig.~\ref{cond1}], when $V_g$ becomes negative, we
observe a narrow Kondo plateau of height $G_0$, and then the conductance drops
suddenly to almost zero at $V_g/U\approx -0.5$, which is exactly the gate potential value 
where QD1 is discharged (see Fig.~\ref{char6}). This drop signals the transitioning 
of QD1 from the Kondo to the IV regime, as described above and in Ref.~\onlinecite{PhysRevB.83.125404}. 
The small peak (of height $0.3~G_0$) observed at $V_g/U\approx-1$ corresponds to the 
discontinuous jump of the renormalized level $\tilde\epsilon_1\approx \varepsilon_1+U^\prime\langle
n_2\rangle$, caused by the discontinuous change in $\langle n_2\rangle$ from $1$ to $0$, leading 
QD1 from a low occupancy ($\langle n_1\rangle \approx 0.0$) IV regime, to a high ($\langle n_1\rangle\approx 2.0$) 
IV regime, completely skipping the Kondo regime (with occupancy $\langle n_1\rangle\approx 1.0$).

When the system reaches the vicinity of its {\it p-h} symmetric point
[$V_g=-(U/2+U^\prime)=-3U/2$], 
there are exactly two electrons in the QDs (one in each) and the Kondo effect is
fully reestablished in QD1 (for $t^\prime=0$), thus the conductance reaches
$G_0$. Due to {\it p-h} symmetry, the other half of the curve can be explained
in a similar way. The (black) $\Circle$ curve in Fig.~\ref{cond1} should be compared 
to the right-side panel of Fig.~3(b) in Ref.~\onlinecite{PhysRevB.83.125404}.

For increasing $t^\prime$ there are mainly two differences in the charge and in
the conductance: i) the abruptness of the charging/discharging and 
 the discontinuities in the conductance are progressively smoothed out,
and ii) the central peak in the conductance (around the {\it p-h}
symmetric point) 
is fully suppressed. For example, for $t^\prime/U=1\times10^{-3}$, 
a rapid drop  in the QD occupations is still clearly visible (as observed on the 
$\square$ (red) curve in Fig.~\ref{char6}), resulting in
a rapid variation in the conductance, which now moves down to a lower value of $V_g$
when compared with the $t^\prime=0$ curve (see $\square$ curve in Fig.~\ref{cond1}). 
Notice that the peak
at $V_g\approx -U$ is enhanced, as $t^\prime$ increases, as now the value of $V_g$ 
in the IV regime corresponds to a higher conductance. For even larger $t^\prime$ 
there is a continuous enhancement of the conductance as $V_g$ approaches $-U$, 
as clearly observed for $t^\prime/U=1.0\times10^{-2}$. The suppression of the
conductance around the {\it p-h} symmetric point for small (but
finite) $t^\prime$ results from destructive Fano-like interference due
to the Kondo resonance in QD2. For large $t^\prime$, on the other hand, results
from the formation of a local singlet due to an antiferromagnetic effective
coupling
($J=4t^{\prime 2}/U$) between the two electrons in the QDs, competing with 
the Kondo singlet formed between the electron in QD1 and the conduction
electrons. This is clearly associated to the TSK effect previously analyzed in
this system. 
\cite{PhysRevB.71.075305}

One should note that there is a striking difference in the effect 
of $t^\prime$ over the conductance for different regions of gate potential, and for different values of $t^\prime$.
 For the smallest value studied ($t^\prime$), 
as mentioned above, the Kondo peak at half-filling is immediately suppressed and this 
can be associated to the TSK effect. On the other hand, the peak around $V_g=-U$
is barely  affected for a small $t^\prime$. Indeed, as the TSK effect
depends on the energy scale $J^\prime$, one expects that it will be more effective at half-filling. 
Higher values of $t^\prime$ will then quickly modify the structures around $V_g=-U/2$: the 
peak at $V_g=-U$ is quickly enhanced and shifted to higher gate voltage values, while 
the discontinuity located (for $t^{\prime}=0$) at $V_g=-U/2$ moves to lower gate potential 
values, extending the Kondo plateau, until the two structures merge and form a 
single Kondo peak centered at $V_g=-U/2$. This occurs because now level $\epsilon_2$ acquires a finite width 
and therefore the discharging of QD1 (and simultaneous charging of QD2) 
is not abrupt anymore and the system smoothly proceeds  from the IV to the Kondo regime.

\subsubsection{Finite-temperature case}

\begin{figure}
\centerline{\resizebox{3.60in}{!}{
\includegraphics{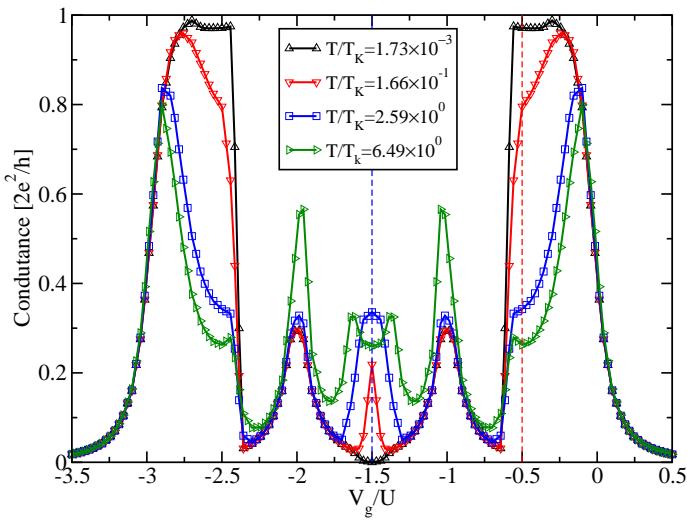}}}
\caption{\label{cond3}(color online)  Conductance as function of $V_g$ for
  $U_2=U_{1}$, $t^\prime/U=1.0\times10^{-3}$
for $T/T_K=7.31\times10^{-10}$ ($\Circle$), $T/T_K=1.08\times10^{-4}$
($\square$), $T/T_K=1.70\times10^{-3}$ ($\Diamond$), $T/T_K=1.66\times10^{-1}$
($\triangle$), $T/T_K=6.49\times10^{0}$ ($\triangleright$).}
\end{figure}

Now we turn our attention to the effect of temperature in the transport
properties  of the system described in Fig.~\ref{model}. In Fig.~\ref{cond3}, we show the conductance
for $U^\prime=U$, $t^\prime/U=1.0\times10^{-3}$, and various values of
temperature. The curves in this figure should  be compared to the
corresponding zero temperature curves in Fig.~\ref{cond1} [(red) $\square$ curve].
Starting with $T/T_K=1.7\times10^{-3}$ [(black) $\bigtriangleup$] we notice a very similar
shape when compared to the $T=0$ results [(red) $\square$] in Fig.~\ref{cond1}, 
except that the two symmetric Kondo plateaus  and the IV peaks at $V_g/U\approx -1$ and 
$V_g/U\approx -2$ are slightly suppressed, which results from the small (but
finite) temperature. Results for smaller temperatures (not shown) interpolate between 
$T=0$ and $T/T_K=1.7 \times 10^{-3}$. 
For $T/T_K=1.66\times10^{-1}$ [(red) $\triangledown$] we observe a clear
suppression of the Kondo plateau while, the IV peaks do not differ
much from those for lower temperatures. This is readily understood since the characteristic
energy of the Kondo state $T_K$ is much smaller than the energy scale associated
to the IV regime, of order $\Gamma_1$ (for details, see Fig.~8 in Ref.~\onlinecite{PhysRevB.83.125404}). 
On the other hand, it is interesting to notice the emergence of a sharp small peak at the {\it p-h} symmetric point for a higher
temperature such as $T/T_K=1.66\times 10^{-1}$ [(red) $\triangledown$]. This peak 
results from a revival of the Kondo peak observed for $t^\prime=0$ in the $T=0$
case, see the corresponding curve in Fig.~\ref{cond1} [(black) $\Circle$]. 
This revival of the Kondo peak results from the progressive
restoration of the Kondo singlet state in QD1 as the local singlet is suppressed
when $T^\prime_K<T<T_K$. 
For $T>T_K$, 
as in the (green) {\large{$\triangleright$}} curve, we observe a suppression of the {\it p-h} Kondo peak, 
similar to a splitting. 
\begin{figure}
\centerline{\resizebox{3.50in}{!}{
\includegraphics{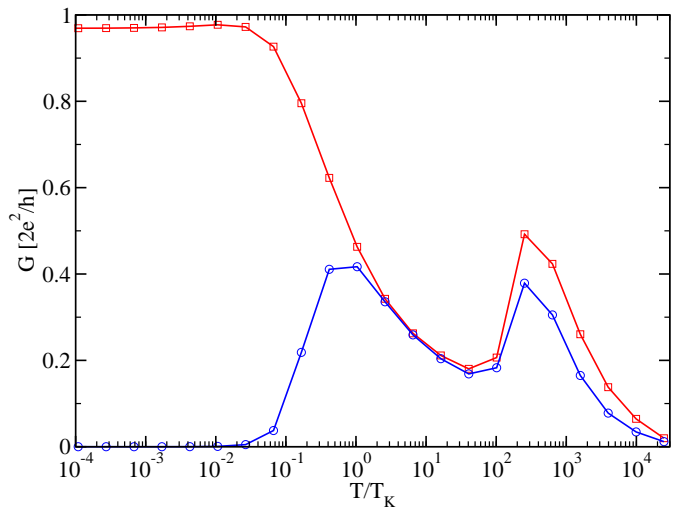}}}
\caption{\label{condT-B} (color online) Conductance as function of temperature
for $t^\prime/U=1.0\times10^{-3}$  for $V_g/U=-1.5$  [(blue) $\Circle$] and
$V_g=-0.5$ [(red) $\square$], along the (blue and red) dashed vertical lines
shown in Fig.~\ref{cond3}.}
\end{figure}

In order to show this effect in more detail, in Fig~\ref{condT-B} we plot 
the temperature variation of the conductance for the same parameters as in
Fig.~\ref{cond3}, for two different values of $V_g$. For $V_g/U=-1.5$
[(blue) $\Circle$ curve], the conductance vanishes when $T\rightarrow 0$, increases for 
$10^{-2}\lesssim T/T_K \lesssim 10^{-1}$, and is suppressed again for $T\gtrsim
T_K$. Note that the initial enhancement of the conductance as the temperature
increases results from the destruction of the Kondo resonance in the QD2,
suppressing the destructive interference between the two paths. Conversely, 
as the temperature keeps increasing, and $T$ exceeds $T_K$, the conductance is suppressed
due to the destruction of the Kondo resonance in QD1. 
Finally, it will rise again and reach a maximum at approximately $T/T_K\sim 10^2\approx \Gamma_1$ due to
charge fluctuations in QD1. For $V_g=-0.5$ [(red) $\square$ curve], i.e., 
at the edge of the right plateau of Fig.~\ref{cond3}], the conductance
starts from $G_0=2e^2/h$ as $T\rightarrow 0$, and is suppressed for $10^{-2}~T_K \lesssim T \lesssim 10~T_K$ due to the destruction of
the Kondo effect in QD1. This is the behavior of the regular Kondo effect in a
single QD. It is interesting to point out that once the temperature is above
$T^\prime_K$, the conductance at both $V_g$ values behave alike.

\begin{figure}
\vskip1cm
\centerline{\resizebox{3.60in}{!}{
\includegraphics{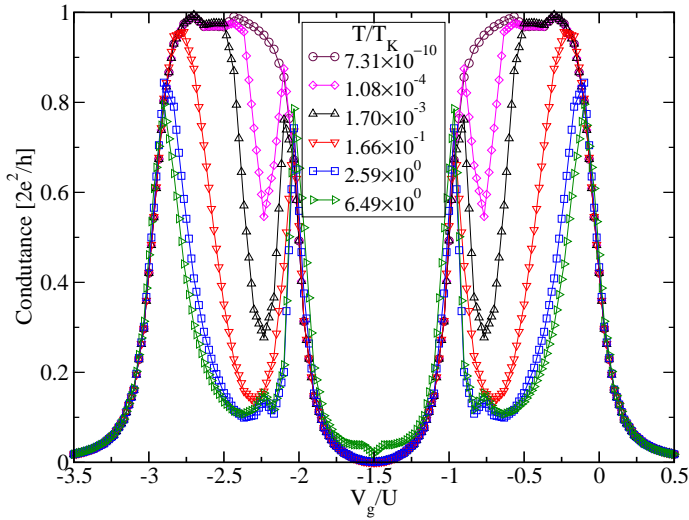}}}
\caption{\label{cond4} (color online)  Conductance as function of $V_g$ and
$t^\prime/U=0.01$
for various
temperatures in the $U_2=U_{1}$, $t^\prime/U=1.0\times10^{-2}$ and
$T/T_K=7.31\times10^{-10}$ ($\Circle$), $T/T_K=1.08\times10^{-4}$ ($\Diamond$),
$T/T_K=1.70\times10^{-3}$ ($\triangle$), $T/T_K=1.66\times10^{-1}$
($\triangledown$), $T/T_K=2.59\times10^{0}$ ($\square$), and
$T/T_K=6.49\times10^{0}$ ($\triangleright$).}
\end{figure}

In Fig.~\ref{cond4}, similarly to Fig.~\ref{cond3}, we show $G~vs.~V_g$, 
for different temperatures, but now for $t^\prime=1\times10^{-2}$, the largest $t^\prime$ value used 
in Fig.~\ref{cond1}. These curves should be compared to the zero-temperature (blue) 
$\triangle$ curve in Fig.~\ref{cond1}. For the case of $T/T_K\sim 10^{-10}$, the
curve has the same shape as the corresponding curve in Fig.~\ref{cond1}. We
notice, however, that by gradually increasing the temperature, both Kondo
plateaus become valleys surrounded by two Coulomb blockade peaks, while the gap
at the {\it p-h} symmetric point remains almost unchanged. 
This is in consistent with what one would expect of the charge transport at high temperature 
for two molecular levels. 

\section{Concluding remarks \label{sec-conclusion}}

We have studied a strongly interacting double dot system arranged in a T-shape
configuration, where the dots are coupled via a tunneling matrix element
and also strong inter-dot Coulomb repulsion. We have presented a detailed
study of the effect of the inter-dot Coulomb interaction on the Kondo
physics of the system. Our numerical analysis reveals interesting crossovers 
 between different mixed valence and Kondo regimes that produce
dramatic changes in the conductance of the system.
These crossovers can be tuned by varying the gate voltage of the QDs, and the
charging/discharging processes produce anomalous peaks in the conductance 
across the system, allowing a clear identification of the various
regimes. 
We show that the inter-dot Coulomb repulsion not only preserves the TSK regime, 
but it also dramatically increases the lowest energy scale of the TSK effect. 
Indeed, $T_K^\prime$ increases exponentially with the inter-dot Coulomb interaction, 
as shown in panels 9a) and (b) of Fig.~\ref{entsus}. We stress that 
this enhancement may allow the experimental observation of this 
so far elusive effect. Finally, by fixing the gate voltage at the {\it p-h} symmetric point and
raising the temperature, the conductance shows clearly the crossover from TSK to
Kondo and then to mixed valence regime (see Fig.~\ref{condT-B}). By contrast,
away from the {\it p-h} symmetry only the regular Kondo regime is
observed.  For temperatures larger than $T_K$ (and large enough values of 
the coupling $t^\prime$ between the QDs) the conductance of the system as a 
function of the gate voltage possesses a four-peak structure, showing clearly
the Coulomb blockade regime of the molecular orbitals of the system. Finally,
we believe that our results will motivate future experimental measurements. 

\acknowledgments
ILF, FMS, and EV acknowledge CNPq, CAPES, and FAPEMIG, the Brazilian agencies, for
financial support. PO acknowledges FONDECYT under grant No. 1100560,  and
GBM acknowledges financial support by the National Science Foundation under
Grant No. DMR-0710529. We also
would like to thank G. A. Lara for valuable discussions.

\bibliography{references}

\begin{thebibliography}{51}
\expandafter\ifx\csname natexlab\endcsname\relax\def\natexlab#1{#1}\fi
\expandafter\ifx\csname bibnamefont\endcsname\relax
  \def\bibnamefont#1{#1}\fi
\expandafter\ifx\csname bibfnamefont\endcsname\relax
  \def\bibfnamefont#1{#1}\fi
\expandafter\ifx\csname citenamefont\endcsname\relax
  \def\citenamefont#1{#1}\fi
\expandafter\ifx\csname url\endcsname\relax
  \def\url#1{\texttt{#1}}\fi
\expandafter\ifx\csname urlprefix\endcsname\relax\def\urlprefix{URL }\fi
\providecommand{\bibinfo}[2]{#2}
\providecommand{\eprint}[2][]{\url{#2}}

\bibitem[{\citenamefont{Reed}(1993)}]{ScientAmerican.268.118}
\bibinfo{author}{\bibfnamefont{M.}~\bibnamefont{Reed}},
  \bibinfo{journal}{Scientific American} \textbf{\bibinfo{volume}{268}},
  \bibinfo{pages}{118} (\bibinfo{year}{1993}).

\bibitem[{\citenamefont{Georges and Meir}(1999)}]{PhysRevLett.82.3508}
\bibinfo{author}{\bibfnamefont{A.}~\bibnamefont{Georges}} \bibnamefont{and}
  \bibinfo{author}{\bibfnamefont{Y.}~\bibnamefont{Meir}},
  \bibinfo{journal}{Phys. Rev. Lett.} \textbf{\bibinfo{volume}{82}},
  \bibinfo{pages}{3508} (\bibinfo{year}{1999}).

\bibitem[{\citenamefont{Sasaki et~al.}(2009)\citenamefont{Sasaki, Tamura,
  Akazaki, and Fujisawa}}]{PhysRevLett.103.266806}
\bibinfo{author}{\bibfnamefont{S.}~\bibnamefont{Sasaki}},
  \bibinfo{author}{\bibfnamefont{H.}~\bibnamefont{Tamura}},
  \bibinfo{author}{\bibfnamefont{T.}~\bibnamefont{Akazaki}}, \bibnamefont{and}
  \bibinfo{author}{\bibfnamefont{T.}~\bibnamefont{Fujisawa}},
  \bibinfo{journal}{Phys. Rev. Lett.} \textbf{\bibinfo{volume}{103}},
  \bibinfo{pages}{266806} (\bibinfo{year}{2009}).

\bibitem[{\citenamefont{Cronenwett et~al.}(1998)\citenamefont{Cronenwett,
  Oosterkamp, and Kouwenhoven}}]{Science.S.1998.540}
\bibinfo{author}{\bibfnamefont{S.~M.} \bibnamefont{Cronenwett}},
  \bibinfo{author}{\bibfnamefont{T.~H.} \bibnamefont{Oosterkamp}},
  \bibnamefont{and} \bibinfo{author}{\bibfnamefont{L.~P.}
  \bibnamefont{Kouwenhoven}}, \bibinfo{journal}{Science}
  \textbf{\bibinfo{volume}{281}}, \bibinfo{pages}{540} (\bibinfo{year}{1998}).

\bibitem[{\citenamefont{Schedelbeck et~al.}(1997)\citenamefont{Schedelbeck,
  Wegscheider, Bichler, and Abstreiter}}]{Science.278.1792}
\bibinfo{author}{\bibfnamefont{G.}~\bibnamefont{Schedelbeck}},
  \bibinfo{author}{\bibfnamefont{W.}~\bibnamefont{Wegscheider}},
  \bibinfo{author}{\bibfnamefont{M.}~\bibnamefont{Bichler}}, \bibnamefont{and}
  \bibinfo{author}{\bibfnamefont{G.}~\bibnamefont{Abstreiter}},
  \bibinfo{journal}{Science} \textbf{\bibinfo{volume}{278}},
  \bibinfo{pages}{1792} (\bibinfo{year}{1997}).

\bibitem[{\citenamefont{Henderson et~al.}(2007)\citenamefont{Henderson, Ramsey,
  del Barco, Mishra, and G.}}]{J.Appl.Phys..101.09E102}
\bibinfo{author}{\bibfnamefont{J.~J.} \bibnamefont{Henderson}},
  \bibinfo{author}{\bibfnamefont{C.~M.} \bibnamefont{Ramsey}},
  \bibinfo{author}{\bibfnamefont{E.}~\bibnamefont{del Barco}},
  \bibinfo{author}{\bibfnamefont{A.}~\bibnamefont{Mishra}}, \bibnamefont{and}
  \bibinfo{author}{\bibfnamefont{C.}~\bibnamefont{G.}}, \bibinfo{journal}{J.
  Appl. Phys.} \textbf{\bibinfo{volume}{101}}, \bibinfo{pages}{09E102}
  (\bibinfo{year}{2007}).

\bibitem[{\citenamefont{Goldhaber-Gordon
  et~al.}(1998{\natexlab{a}})\citenamefont{Goldhaber-Gordon, Shtrikman, Mahalu,
  Abusch-Magder, Meirav, and Kastner}}]{Nature.391.156}
\bibinfo{author}{\bibfnamefont{D.}~\bibnamefont{Goldhaber-Gordon}},
  \bibinfo{author}{\bibfnamefont{H.}~\bibnamefont{Shtrikman}},
  \bibinfo{author}{\bibfnamefont{D.}~\bibnamefont{Mahalu}},
  \bibinfo{author}{\bibfnamefont{D.}~\bibnamefont{Abusch-Magder}},
  \bibinfo{author}{\bibfnamefont{U.}~\bibnamefont{Meirav}}, \bibnamefont{and}
  \bibinfo{author}{\bibfnamefont{M.~A.} \bibnamefont{Kastner}},
  \bibinfo{journal}{Nature} \textbf{\bibinfo{volume}{391}},
  \bibinfo{pages}{156} (\bibinfo{year}{1998}{\natexlab{a}}).

\bibitem[{\citenamefont{Kouwenhoven and Marcus}(1998)}]{Phys.World.11.35}
\bibinfo{author}{\bibfnamefont{L.}~\bibnamefont{Kouwenhoven}} \bibnamefont{and}
  \bibinfo{author}{\bibfnamefont{C.}~\bibnamefont{Marcus}},
  \bibinfo{journal}{Phys. World} \textbf{\bibinfo{volume}{11}},
  \bibinfo{pages}{35} (\bibinfo{year}{1998}).

\bibitem[{\citenamefont{Rontani et~al.}(1999)\citenamefont{Rontani, Rossi,
  Manghi, and Molinari}}]{PhysRevB.59.10165}
\bibinfo{author}{\bibfnamefont{M.}~\bibnamefont{Rontani}},
  \bibinfo{author}{\bibfnamefont{F.}~\bibnamefont{Rossi}},
  \bibinfo{author}{\bibfnamefont{F.}~\bibnamefont{Manghi}}, \bibnamefont{and}
  \bibinfo{author}{\bibfnamefont{E.}~\bibnamefont{Molinari}},
  \bibinfo{journal}{Phys. Rev. B} \textbf{\bibinfo{volume}{59}},
  \bibinfo{pages}{10165} (\bibinfo{year}{1999}).

\bibitem[{\citenamefont{Brouwer and Aleiner}(1999)}]{PhysRevLett.82.390}
\bibinfo{author}{\bibfnamefont{P.~W.} \bibnamefont{Brouwer}} \bibnamefont{and}
  \bibinfo{author}{\bibfnamefont{I.~L.} \bibnamefont{Aleiner}},
  \bibinfo{journal}{Phys. Rev. Lett.} \textbf{\bibinfo{volume}{82}},
  \bibinfo{pages}{390} (\bibinfo{year}{1999}).

\bibitem[{\citenamefont{Vorojtsov and Baranger}(2005)}]{PhysRevB.72.165349}
\bibinfo{author}{\bibfnamefont{S.}~\bibnamefont{Vorojtsov}} \bibnamefont{and}
  \bibinfo{author}{\bibfnamefont{H.~U.} \bibnamefont{Baranger}},
  \bibinfo{journal}{Phys. Rev. B} \textbf{\bibinfo{volume}{72}},
  \bibinfo{pages}{165349} (\bibinfo{year}{2005}).

\bibitem[{\citenamefont{Goldhaber-Gordon
  et~al.}(1998{\natexlab{b}})\citenamefont{Goldhaber-Gordon, G\"ores, Kastner,
  Shtrikman, Mahalu, and Meirav}}]{PhysRevLett.81.5225}
\bibinfo{author}{\bibfnamefont{D.}~\bibnamefont{Goldhaber-Gordon}},
  \bibinfo{author}{\bibfnamefont{J.}~\bibnamefont{G\"ores}},
  \bibinfo{author}{\bibfnamefont{M.~A.} \bibnamefont{Kastner}},
  \bibinfo{author}{\bibfnamefont{H.}~\bibnamefont{Shtrikman}},
  \bibinfo{author}{\bibfnamefont{D.}~\bibnamefont{Mahalu}}, \bibnamefont{and}
  \bibinfo{author}{\bibfnamefont{U.}~\bibnamefont{Meirav}},
  \bibinfo{journal}{Phys. Rev. Lett.} \textbf{\bibinfo{volume}{81}},
  \bibinfo{pages}{5225} (\bibinfo{year}{1998}{\natexlab{b}}).

\bibitem[{\citenamefont{Jeong et~al.}(2001)\citenamefont{Jeong, Chang, and
  Melloch}}]{Science.293.2221}
\bibinfo{author}{\bibfnamefont{H.}~\bibnamefont{Jeong}},
  \bibinfo{author}{\bibfnamefont{A.~M.} \bibnamefont{Chang}}, \bibnamefont{and}
  \bibinfo{author}{\bibfnamefont{M.~R.} \bibnamefont{Melloch}},
  \bibinfo{journal}{Science} \textbf{\bibinfo{volume}{293}},
  \bibinfo{pages}{2221} (\bibinfo{year}{2001}).

\bibitem[{\citenamefont{van~der Wiel et~al.}(2000)\citenamefont{van~der Wiel,
  Franceschi, Fujisawa, Elzerman, Tarucha, and Kouwenhoven}}]{Science.289.2105}
\bibinfo{author}{\bibfnamefont{W.~G.} \bibnamefont{van~der Wiel}},
  \bibinfo{author}{\bibfnamefont{S.~D.} \bibnamefont{Franceschi}},
  \bibinfo{author}{\bibfnamefont{T.}~\bibnamefont{Fujisawa}},
  \bibinfo{author}{\bibfnamefont{J.~M.} \bibnamefont{Elzerman}},
  \bibinfo{author}{\bibfnamefont{S.}~\bibnamefont{Tarucha}}, \bibnamefont{and}
  \bibinfo{author}{\bibfnamefont{L.~P.} \bibnamefont{Kouwenhoven}},
  \bibinfo{journal}{Science} \textbf{\bibinfo{volume}{289}},
  \bibinfo{pages}{2105} (\bibinfo{year}{2000}).

\bibitem[{\citenamefont{Feng~Chi and
  Zheng}(2008)}]{NanoscaleResearchLetters.3.343}
\bibinfo{author}{\bibfnamefont{X.~Y.} \bibnamefont{Feng~Chi}} \bibnamefont{and}
  \bibinfo{author}{\bibfnamefont{J.}~\bibnamefont{Zheng}},
  \bibinfo{journal}{Nanoscale Research Letters} \textbf{\bibinfo{volume}{3}},
  \bibinfo{pages}{343} (\bibinfo{year}{2008}).

\bibitem[{\citenamefont{\ifmmode~\check{Z}\else \v{Z}\fi{}itko and
  Bon\ifmmode~\check{c}\else \v{c}\fi{}a}(2006)}]{PhysRevB.73.035332}
\bibinfo{author}{\bibfnamefont{R.}~\bibnamefont{\ifmmode~\check{Z}\else
  \v{Z}\fi{}itko}} \bibnamefont{and}
  \bibinfo{author}{\bibfnamefont{J.}~\bibnamefont{Bon\ifmmode~\check{c}\else
  \v{c}\fi{}a}}, \bibinfo{journal}{Phys. Rev. B} \textbf{\bibinfo{volume}{73}},
  \bibinfo{pages}{035332} (\bibinfo{year}{2006}).

\bibitem[{\citenamefont{Dong and Lei}(2002)}]{PhysRevB.65.241304}
\bibinfo{author}{\bibfnamefont{B.}~\bibnamefont{Dong}} \bibnamefont{and}
  \bibinfo{author}{\bibfnamefont{X.~L.} \bibnamefont{Lei}},
  \bibinfo{journal}{Phys. Rev. B} \textbf{\bibinfo{volume}{65}},
  \bibinfo{pages}{241304} (\bibinfo{year}{2002}).

\bibitem[{\citenamefont{Vernek et~al.}(2006)\citenamefont{Vernek, Sandler,
  Ulloa, and Anda}}]{PhysicaE.34.608}
\bibinfo{author}{\bibfnamefont{E.}~\bibnamefont{Vernek}},
  \bibinfo{author}{\bibfnamefont{N.}~\bibnamefont{Sandler}},
  \bibinfo{author}{\bibfnamefont{S.~E.} \bibnamefont{Ulloa}}, \bibnamefont{and}
  \bibinfo{author}{\bibfnamefont{E.~V.} \bibnamefont{Anda}},
  \bibinfo{journal}{Physica E: Low-dimensional Systems and Nanostructures}
  \textbf{\bibinfo{volume}{34}}, \bibinfo{pages}{608} (\bibinfo{year}{2006}).

\bibitem[{\citenamefont{Anda et~al.}(2008)\citenamefont{Anda, Chiappe,
  B\"usser, Davidovich, Martins, Heidrich-Meisner, and
  Dagotto}}]{PhysRevB.78.085308}
\bibinfo{author}{\bibfnamefont{E.~V.} \bibnamefont{Anda}},
  \bibinfo{author}{\bibfnamefont{G.}~\bibnamefont{Chiappe}},
  \bibinfo{author}{\bibfnamefont{C.~A.} \bibnamefont{B\"usser}},
  \bibinfo{author}{\bibfnamefont{M.~A.} \bibnamefont{Davidovich}},
  \bibinfo{author}{\bibfnamefont{G.~B.} \bibnamefont{Martins}},
  \bibinfo{author}{\bibfnamefont{F.}~\bibnamefont{Heidrich-Meisner}},
  \bibnamefont{and} \bibinfo{author}{\bibfnamefont{E.}~\bibnamefont{Dagotto}},
  \bibinfo{journal}{Phys. Rev. B} \textbf{\bibinfo{volume}{78}},
  \bibinfo{pages}{085308} (\bibinfo{year}{2008}).

\bibitem[{\citenamefont{Schuster et~al.}(1997)\citenamefont{Schuster, Buks,
  Heiblum, Mahalu, Umansky, and Shtrikman}}]{Nature.385.417}
\bibinfo{author}{\bibfnamefont{R.}~\bibnamefont{Schuster}},
  \bibinfo{author}{\bibfnamefont{E.}~\bibnamefont{Buks}},
  \bibinfo{author}{\bibfnamefont{M.}~\bibnamefont{Heiblum}},
  \bibinfo{author}{\bibfnamefont{D.}~\bibnamefont{Mahalu}},
  \bibinfo{author}{\bibfnamefont{V.}~\bibnamefont{Umansky}}, \bibnamefont{and}
  \bibinfo{author}{\bibfnamefont{H.}~\bibnamefont{Shtrikman}},
  \bibinfo{journal}{Nature} \textbf{\bibinfo{volume}{385}},
  \bibinfo{pages}{417} (\bibinfo{year}{1997}).

\bibitem[{\citenamefont{Weymann}(2007)}]{PhysRevB.75.195339}
\bibinfo{author}{\bibfnamefont{I.}~\bibnamefont{Weymann}},
  \bibinfo{journal}{Phys. Rev. B} \textbf{\bibinfo{volume}{75}},
  \bibinfo{pages}{195339} (\bibinfo{year}{2007}).

\bibitem[{\citenamefont{Cornaglia and Grempel}(2005)}]{PhysRevB.71.075305}
\bibinfo{author}{\bibfnamefont{P.~S.} \bibnamefont{Cornaglia}}
  \bibnamefont{and} \bibinfo{author}{\bibfnamefont{D.~R.}
  \bibnamefont{Grempel}}, \bibinfo{journal}{Phys. Rev. B}
  \textbf{\bibinfo{volume}{71}}, \bibinfo{pages}{075305}
  (\bibinfo{year}{2005}).

\bibitem[{\citenamefont{Dias~da Silva et~al.}(2006)\citenamefont{Dias~da Silva,
  Sandler, Ingersent, and Ulloa}}]{PhysRevLett.97.096603}
\bibinfo{author}{\bibfnamefont{L.~G. G.~V.} \bibnamefont{Dias~da Silva}},
  \bibinfo{author}{\bibfnamefont{N.~P.} \bibnamefont{Sandler}},
  \bibinfo{author}{\bibfnamefont{K.}~\bibnamefont{Ingersent}},
  \bibnamefont{and} \bibinfo{author}{\bibfnamefont{S.~E.} \bibnamefont{Ulloa}},
  \bibinfo{journal}{Phys. Rev. Lett.} \textbf{\bibinfo{volume}{97}},
  \bibinfo{pages}{096603} (\bibinfo{year}{2006}).

\bibitem[{\citenamefont{Climente et~al.}(2007)\citenamefont{Climente, Bertoni,
  Goldoni, Rontani, and Molinari}}]{PhysRevB.76.085305}
\bibinfo{author}{\bibfnamefont{J.~I.} \bibnamefont{Climente}},
  \bibinfo{author}{\bibfnamefont{A.}~\bibnamefont{Bertoni}},
  \bibinfo{author}{\bibfnamefont{G.}~\bibnamefont{Goldoni}},
  \bibinfo{author}{\bibfnamefont{M.}~\bibnamefont{Rontani}}, \bibnamefont{and}
  \bibinfo{author}{\bibfnamefont{E.}~\bibnamefont{Molinari}},
  \bibinfo{journal}{Phys. Rev. B} \textbf{\bibinfo{volume}{76}},
  \bibinfo{pages}{085305} (\bibinfo{year}{2007}).

\bibitem[{\citenamefont{\ifmmode~\check{Z}\else \v{Z}\fi{}itko and
  Bon\ifmmode~\check{c}\else \v{c}\fi{}a}(2007)}]{PhysRevLett.98.047203}
\bibinfo{author}{\bibfnamefont{R.}~\bibnamefont{\ifmmode~\check{Z}\else
  \v{Z}\fi{}itko}} \bibnamefont{and}
  \bibinfo{author}{\bibfnamefont{J.}~\bibnamefont{Bon\ifmmode~\check{c}\else
  \v{c}\fi{}a}}, \bibinfo{journal}{Phys. Rev. Lett.}
  \textbf{\bibinfo{volume}{98}}, \bibinfo{pages}{047203}
  (\bibinfo{year}{2007}).

\bibitem[{\citenamefont{Vernek et~al.}(2009)\citenamefont{Vernek, B\"usser,
  Martins, Anda, Sandler, and Ulloa}}]{PhysRevB.80.035119}
\bibinfo{author}{\bibfnamefont{E.}~\bibnamefont{Vernek}},
  \bibinfo{author}{\bibfnamefont{C.~A.} \bibnamefont{B\"usser}},
  \bibinfo{author}{\bibfnamefont{G.~B.} \bibnamefont{Martins}},
  \bibinfo{author}{\bibfnamefont{E.~V.} \bibnamefont{Anda}},
  \bibinfo{author}{\bibfnamefont{N.}~\bibnamefont{Sandler}}, \bibnamefont{and}
  \bibinfo{author}{\bibfnamefont{S.~E.} \bibnamefont{Ulloa}},
  \bibinfo{journal}{Phys. Rev. B} \textbf{\bibinfo{volume}{80}},
  \bibinfo{pages}{035119} (\bibinfo{year}{2009}).

\bibitem[{\citenamefont{Trocha and Barna\ifmmode~\acute{s}\else
  \'{s}\fi{}}(2008)}]{PhysRevB.78.075424}
\bibinfo{author}{\bibfnamefont{P.}~\bibnamefont{Trocha}} \bibnamefont{and}
  \bibinfo{author}{\bibfnamefont{J.}~\bibnamefont{Barna\ifmmode~\acute{s}\else
  \'{s}\fi{}}}, \bibinfo{journal}{Phys. Rev. B} \textbf{\bibinfo{volume}{78}},
  \bibinfo{pages}{075424} (\bibinfo{year}{2008}).

\bibitem[{\citenamefont{Pustilnik et~al.}(2003)\citenamefont{Pustilnik,
  Glazman, and Hofstetter}}]{PhysRevB.68.161303}
\bibinfo{author}{\bibfnamefont{M.}~\bibnamefont{Pustilnik}},
  \bibinfo{author}{\bibfnamefont{L.~I.} \bibnamefont{Glazman}},
  \bibnamefont{and}
  \bibinfo{author}{\bibfnamefont{W.}~\bibnamefont{Hofstetter}},
  \bibinfo{journal}{Phys. Rev. B} \textbf{\bibinfo{volume}{68}},
  \bibinfo{pages}{161303} (\bibinfo{year}{2003}).

\bibitem[{\citenamefont{Mitchell and Logan}(2010)}]{PhysRevB.81.075126}
\bibinfo{author}{\bibfnamefont{A.~K.} \bibnamefont{Mitchell}} \bibnamefont{and}
  \bibinfo{author}{\bibfnamefont{D.~E.} \bibnamefont{Logan}},
  \bibinfo{journal}{Phys. Rev. B} \textbf{\bibinfo{volume}{81}},
  \bibinfo{pages}{075126} (\bibinfo{year}{2010}).

\bibitem[{\citenamefont{Rogge and Haug}(2008)}]{PhysRevB.77.193306}
\bibinfo{author}{\bibfnamefont{M.~C.} \bibnamefont{Rogge}} \bibnamefont{and}
  \bibinfo{author}{\bibfnamefont{R.~J.} \bibnamefont{Haug}},
  \bibinfo{journal}{Phys. Rev. B} \textbf{\bibinfo{volume}{77}},
  \bibinfo{pages}{193306} (\bibinfo{year}{2008}).

\bibitem[{\citenamefont{Chiappe et~al.}(2010)\citenamefont{Chiappe, Anda,
  Costa~Ribeiro, and Louis}}]{PhysRevB.81.041310}
\bibinfo{author}{\bibfnamefont{G.}~\bibnamefont{Chiappe}},
  \bibinfo{author}{\bibfnamefont{E.~V.} \bibnamefont{Anda}},
  \bibinfo{author}{\bibfnamefont{L.}~\bibnamefont{Costa~Ribeiro}},
  \bibnamefont{and} \bibinfo{author}{\bibfnamefont{E.}~\bibnamefont{Louis}},
  \bibinfo{journal}{Phys. Rev. B} \textbf{\bibinfo{volume}{81}},
  \bibinfo{pages}{041310} (\bibinfo{year}{2010}).

\bibitem[{\citenamefont{Chan et~al.}(2002)\citenamefont{Chan, Westervelt,
  Maranowski, and Gossard}}]{Appl.Phys.Lett..80.1818}
\bibinfo{author}{\bibfnamefont{I.}~\bibnamefont{Chan}},
  \bibinfo{author}{\bibfnamefont{R.~M.} \bibnamefont{Westervelt}},
  \bibinfo{author}{\bibfnamefont{K.~D.} \bibnamefont{Maranowski}},
  \bibnamefont{and} \bibinfo{author}{\bibfnamefont{A.~C.}
  \bibnamefont{Gossard}}, \bibinfo{journal}{Appl. Phys. Lett.}
  \textbf{\bibinfo{volume}{80}}, \bibinfo{pages}{1818} (\bibinfo{year}{2002}).

\bibitem[{\citenamefont{Holleitner et~al.}(2003)\citenamefont{Holleitner,
  Blick, , and Eberl}}]{Appl.Phys.Lett..82.1887}
\bibinfo{author}{\bibfnamefont{A.~W.} \bibnamefont{Holleitner}},
  \bibinfo{author}{\bibfnamefont{R.~H.} \bibnamefont{Blick}}, ,
  \bibnamefont{and} \bibinfo{author}{\bibfnamefont{K.}~\bibnamefont{Eberl}},
  \bibinfo{journal}{Appl. Phys. Lett.} \textbf{\bibinfo{volume}{82}},
  \bibinfo{pages}{1887} (\bibinfo{year}{2003}).

\bibitem[{\citenamefont{McClure et~al.}(2007)\citenamefont{McClure, DiCarlo,
  Zhang, Engel, Marcus, Hanson, and Gossard}}]{PhysRevLett.98.056801}
\bibinfo{author}{\bibfnamefont{D.~T.} \bibnamefont{McClure}},
  \bibinfo{author}{\bibfnamefont{L.}~\bibnamefont{DiCarlo}},
  \bibinfo{author}{\bibfnamefont{Y.}~\bibnamefont{Zhang}},
  \bibinfo{author}{\bibfnamefont{H.-A.} \bibnamefont{Engel}},
  \bibinfo{author}{\bibfnamefont{C.~M.} \bibnamefont{Marcus}},
  \bibinfo{author}{\bibfnamefont{M.~P.} \bibnamefont{Hanson}},
  \bibnamefont{and} \bibinfo{author}{\bibfnamefont{A.~C.}
  \bibnamefont{Gossard}}, \bibinfo{journal}{Phys. Rev. Lett.}
  \textbf{\bibinfo{volume}{98}}, \bibinfo{pages}{056801}
  (\bibinfo{year}{2007}).

\bibitem[{\citenamefont{H\"ubel et~al.}(2008)\citenamefont{H\"ubel, Held, Weis,
  and v.~Klitzing}}]{PhysRevLett.101.186804}
\bibinfo{author}{\bibfnamefont{A.}~\bibnamefont{H\"ubel}},
  \bibinfo{author}{\bibfnamefont{K.}~\bibnamefont{Held}},
  \bibinfo{author}{\bibfnamefont{J.}~\bibnamefont{Weis}}, \bibnamefont{and}
  \bibinfo{author}{\bibfnamefont{K.}~\bibnamefont{v.~Klitzing}},
  \bibinfo{journal}{Phys. Rev. Lett.} \textbf{\bibinfo{volume}{101}},
  \bibinfo{pages}{186804} (\bibinfo{year}{2008}).

\bibitem[{\citenamefont{Hatano et~al.}(2010)\citenamefont{Hatano, Amaha, Kubo,
  Teraoka, Tokura, Gupta, Austing, and
  S.}}]{arXiv:1008.0071v1.cond-mat.mes-hall..}
\bibinfo{author}{\bibfnamefont{T.}~\bibnamefont{Hatano}},
  \bibinfo{author}{\bibfnamefont{S.}~\bibnamefont{Amaha}},
  \bibinfo{author}{\bibfnamefont{T.}~\bibnamefont{Kubo}},
  \bibinfo{author}{\bibfnamefont{S.}~\bibnamefont{Teraoka}},
  \bibinfo{author}{\bibfnamefont{Y.}~\bibnamefont{Tokura}},
  \bibinfo{author}{\bibfnamefont{J.~A.} \bibnamefont{Gupta}},
  \bibinfo{author}{\bibfnamefont{D.~G.} \bibnamefont{Austing}},
  \bibnamefont{and} \bibinfo{author}{\bibfnamefont{T.}~\bibnamefont{S.}},
  \bibinfo{journal}{arXiv:1008.0071v1 [cond-mat.mes-hall]}
  (\bibinfo{year}{2010}).

\bibitem[{\citenamefont{Andergassen et~al.}(2008)\citenamefont{Andergassen,
  Simon, Florens, and Feinberg}}]{PhysRevB.77.045309}
\bibinfo{author}{\bibfnamefont{S.}~\bibnamefont{Andergassen}},
  \bibinfo{author}{\bibfnamefont{P.}~\bibnamefont{Simon}},
  \bibinfo{author}{\bibfnamefont{S.}~\bibnamefont{Florens}}, \bibnamefont{and}
  \bibinfo{author}{\bibfnamefont{D.}~\bibnamefont{Feinberg}},
  \bibinfo{journal}{Phys. Rev. B} \textbf{\bibinfo{volume}{77}},
  \bibinfo{pages}{045309} (\bibinfo{year}{2008}).

\bibitem[{\citenamefont{Chen et~al.}(1996)\citenamefont{Chen, Korotkov, and
  Likharev}}]{Appl.Phys.Lett..68.1954}
\bibinfo{author}{\bibfnamefont{R.~H.} \bibnamefont{Chen}},
  \bibinfo{author}{\bibfnamefont{A.~N.} \bibnamefont{Korotkov}},
  \bibnamefont{and} \bibinfo{author}{\bibfnamefont{K.~K.}
  \bibnamefont{Likharev}}, \bibinfo{journal}{Appl. Phys. Lett.}
  \textbf{\bibinfo{volume}{68}}, \bibinfo{pages}{1954} (\bibinfo{year}{1996}).

\bibitem[{\citenamefont{Ivanov et~al.}(1994)\citenamefont{Ivanov, Valtchinov,
  and Wille}}]{PhysRevB.50.4917}
\bibinfo{author}{\bibfnamefont{T.}~\bibnamefont{Ivanov}},
  \bibinfo{author}{\bibfnamefont{V.}~\bibnamefont{Valtchinov}},
  \bibnamefont{and} \bibinfo{author}{\bibfnamefont{L.~T.} \bibnamefont{Wille}},
  \bibinfo{journal}{Phys. Rev. B} \textbf{\bibinfo{volume}{50}},
  \bibinfo{pages}{4917} (\bibinfo{year}{1994}).

\bibitem[{\citenamefont{Kashcheyevs et~al.}(2009)\citenamefont{Kashcheyevs,
  Karrasch, Hecht, Weichselbaum, Meden, and Schiller}}]{PhysRevLett.102.136805}
\bibinfo{author}{\bibfnamefont{V.}~\bibnamefont{Kashcheyevs}},
  \bibinfo{author}{\bibfnamefont{C.}~\bibnamefont{Karrasch}},
  \bibinfo{author}{\bibfnamefont{T.}~\bibnamefont{Hecht}},
  \bibinfo{author}{\bibfnamefont{A.}~\bibnamefont{Weichselbaum}},
  \bibinfo{author}{\bibfnamefont{V.}~\bibnamefont{Meden}}, \bibnamefont{and}
  \bibinfo{author}{\bibfnamefont{A.}~\bibnamefont{Schiller}},
  \bibinfo{journal}{Phys. Rev. Lett.} \textbf{\bibinfo{volume}{102}},
  \bibinfo{pages}{136805} (\bibinfo{year}{2009}).

\bibitem[{\citenamefont{Moldoveanu et~al.}(2010)\citenamefont{Moldoveanu,
  Manolescu, and Gudmundsson}}]{PhysRevB.82.085311}
\bibinfo{author}{\bibfnamefont{V.}~\bibnamefont{Moldoveanu}},
  \bibinfo{author}{\bibfnamefont{A.}~\bibnamefont{Manolescu}},
  \bibnamefont{and}
  \bibinfo{author}{\bibfnamefont{V.}~\bibnamefont{Gudmundsson}},
  \bibinfo{journal}{Phys. Rev. B} \textbf{\bibinfo{volume}{82}},
  \bibinfo{pages}{085311} (\bibinfo{year}{2010}).

\bibitem[{\citenamefont{Galpin et~al.}(2006)\citenamefont{Galpin, Logan, and
  Krishnamurthy}}]{J.Phys.:Condens.Matter.18.6545}
\bibinfo{author}{\bibfnamefont{M.~R.} \bibnamefont{Galpin}},
  \bibinfo{author}{\bibfnamefont{D.~E.} \bibnamefont{Logan}}, \bibnamefont{and}
  \bibinfo{author}{\bibfnamefont{H.~R.} \bibnamefont{Krishnamurthy}},
  \bibinfo{journal}{J. Phys.: Condens. Matter} \textbf{\bibinfo{volume}{18}},
  \bibinfo{pages}{6545} (\bibinfo{year}{2006}).

\bibitem[{\citenamefont{Granger et~al.}(2005)\citenamefont{Granger, Kastner,
  Radu, Hanson, and Gossard}}]{PhysRevB.72.165309}
\bibinfo{author}{\bibfnamefont{G.}~\bibnamefont{Granger}},
  \bibinfo{author}{\bibfnamefont{M.~A.} \bibnamefont{Kastner}},
  \bibinfo{author}{\bibfnamefont{I.}~\bibnamefont{Radu}},
  \bibinfo{author}{\bibfnamefont{M.~P.} \bibnamefont{Hanson}},
  \bibnamefont{and} \bibinfo{author}{\bibfnamefont{A.~C.}
  \bibnamefont{Gossard}}, \bibinfo{journal}{Phys. Rev. B}
  \textbf{\bibinfo{volume}{72}}, \bibinfo{pages}{165309}
  (\bibinfo{year}{2005}).

\bibitem[{\citenamefont{Chung et~al.}(2008)\citenamefont{Chung, Zarand, and
  W\"olfle}}]{PhysRevB.77.035120}
\bibinfo{author}{\bibfnamefont{C.-H.} \bibnamefont{Chung}},
  \bibinfo{author}{\bibfnamefont{G.}~\bibnamefont{Zarand}}, \bibnamefont{and}
  \bibinfo{author}{\bibfnamefont{P.}~\bibnamefont{W\"olfle}},
  \bibinfo{journal}{Phys. Rev. B} \textbf{\bibinfo{volume}{77}},
  \bibinfo{pages}{035120} (\bibinfo{year}{2008}).

\bibitem[{\citenamefont{\ifmmode~\check{Z}\else
  \v{Z}\fi{}itko}(2010)}]{PhysRevB.81.115316}
\bibinfo{author}{\bibfnamefont{R.}~\bibnamefont{\ifmmode~\check{Z}\else
  \v{Z}\fi{}itko}}, \bibinfo{journal}{Phys. Rev. B}
  \textbf{\bibinfo{volume}{81}}, \bibinfo{pages}{115316}
  (\bibinfo{year}{2010}).

\bibitem[{\citenamefont{Wilson}(1975)}]{RevModPhys.47.773}
\bibinfo{author}{\bibfnamefont{K.~G.} \bibnamefont{Wilson}},
  \bibinfo{journal}{Rev. Mod. Phys.} \textbf{\bibinfo{volume}{47}},
  \bibinfo{pages}{773} (\bibinfo{year}{1975}).

\bibitem[{\citenamefont{Krishna-murthy
  et~al.}(1980{\natexlab{a}})\citenamefont{Krishna-murthy, Wilkins, and
  Wilson}}]{PhysRevB.21.1003}
\bibinfo{author}{\bibfnamefont{H.~R.} \bibnamefont{Krishna-murthy}},
  \bibinfo{author}{\bibfnamefont{J.~W.} \bibnamefont{Wilkins}},
  \bibnamefont{and} \bibinfo{author}{\bibfnamefont{K.~G.}
  \bibnamefont{Wilson}}, \bibinfo{journal}{Phys. Rev. B}
  \textbf{\bibinfo{volume}{21}}, \bibinfo{pages}{1003}
  (\bibinfo{year}{1980}{\natexlab{a}}).

\bibitem[{\citenamefont{Krishna-murthy
  et~al.}(1980{\natexlab{b}})\citenamefont{Krishna-murthy, Wilkins, and
  Wilson}}]{PhysRevB.21.1044}
\bibinfo{author}{\bibfnamefont{H.~R.} \bibnamefont{Krishna-murthy}},
  \bibinfo{author}{\bibfnamefont{J.~W.} \bibnamefont{Wilkins}},
  \bibnamefont{and} \bibinfo{author}{\bibfnamefont{K.~G.}
  \bibnamefont{Wilson}}, \bibinfo{journal}{Phys. Rev. B}
  \textbf{\bibinfo{volume}{21}}, \bibinfo{pages}{1044}
  (\bibinfo{year}{1980}{\natexlab{b}}).

\bibitem[{\citenamefont{Bulla et~al.}(2001)\citenamefont{Bulla, Costi, and
  Vollhardt}}]{PhysRevB.64.045103}
\bibinfo{author}{\bibfnamefont{R.}~\bibnamefont{Bulla}},
  \bibinfo{author}{\bibfnamefont{T.~A.} \bibnamefont{Costi}}, \bibnamefont{and}
  \bibinfo{author}{\bibfnamefont{D.}~\bibnamefont{Vollhardt}},
  \bibinfo{journal}{Phys. Rev. B} \textbf{\bibinfo{volume}{64}},
  \bibinfo{pages}{045103} (\bibinfo{year}{2001}).

\bibitem[{\citenamefont{B\"usser et~al.}(2011)\citenamefont{B\"usser, Vernek,
  Orellana, Lara, Kim, Feiguin, Anda, and Martins}}]{PhysRevB.83.125404}
\bibinfo{author}{\bibfnamefont{C.~A.} \bibnamefont{B\"usser}},
  \bibinfo{author}{\bibfnamefont{E.}~\bibnamefont{Vernek}},
  \bibinfo{author}{\bibfnamefont{P.}~\bibnamefont{Orellana}},
  \bibinfo{author}{\bibfnamefont{G.~A.} \bibnamefont{Lara}},
  \bibinfo{author}{\bibfnamefont{E.~H.} \bibnamefont{Kim}},
  \bibinfo{author}{\bibfnamefont{A.~E.} \bibnamefont{Feiguin}},
  \bibinfo{author}{\bibfnamefont{E.~V.} \bibnamefont{Anda}}, \bibnamefont{and}
  \bibinfo{author}{\bibfnamefont{G.~B.} \bibnamefont{Martins}},
  \bibinfo{journal}{Phys. Rev. B} \textbf{\bibinfo{volume}{83}},
  \bibinfo{pages}{125404} (\bibinfo{year}{2011}).

\bibitem[{\citenamefont{Pemberton-Ross
  et~al.}(2010)\citenamefont{Pemberton-Ross, Kay, and
  Schirmer}}]{PhysRevA.82.042322}
\bibinfo{author}{\bibfnamefont{P.~J.} \bibnamefont{Pemberton-Ross}},
  \bibinfo{author}{\bibfnamefont{A.}~\bibnamefont{Kay}}, \bibnamefont{and}
  \bibinfo{author}{\bibfnamefont{S.~G.} \bibnamefont{Schirmer}},
  \bibinfo{journal}{Phys. Rev. A} \textbf{\bibinfo{volume}{82}},
  \bibinfo{pages}{042322} (\bibinfo{year}{2010}).

\end{thebibliography}
\end{document}